# Multi-microjoule GaSe-based mid-infrared optical parametric amplifier with an ultra-broad idler spectrum covering 4.2-16 µm

Kun Liu[1,2], Houkun Liang[1,*], Lifeng Wang[1], Shizhen Qu[1,2], Tino Lang[3], Hao Li[1], Qi Jie Wang[2,*], Ying Zhang[1]

1 Precision Measurements Group, Singapore Institute of Manufacturing Technology, 2 Fusionopolis Way, 138634, Singapore
2 School of Electrical & Electronic Engineering & The Photonics Institute, Nanyang Technological University 639798, Singapore
3 Center for Free-Electron Laser Science, DESY, Notkestraße 85, 22607 Hamburg, Germany
*Corresponding author: hkliang@simtech.a-star.edu.sg, qjwang@ntu.edu.sg

**We report a multi-microjoule, ultra-broadband mid-infrared optical parametric amplifier based on a GaSe nonlinear crystal pumped at ~2 µm. The generated idler pulse has a flat spectrum spanning from 4.5 to 13.3 µm at -3 dB and 4.2 to 16 µm in the full spectral range, with a central wavelength of 8.8 µm. The proposed scheme supports a sub-cycle Fourier-transform-limited pulse width. A (2+1)-dimensional numerical simulation is employed to reproduce the obtained idler spectrum. To our best knowledge, this is the broadest -3 dB spectrum ever obtained by optical parametric amplifiers in this spectral region. The idler pulse energy is ~3.4 µJ with a conversion efficiency of ~2% from the ~2 µm pump to the idler pulse.**

http://dx.doi.org/10.1364/OL.99.099999

High-energy ultra-broadband few-cycle mid-infrared (mid-IR) sources have attracted extensive attention for their applications in molecular spectroscopy [1-4] and strong-field physics [5-14]. In recent years, the development of such sources has been extended to >6 µm long-wavelength region. In the perspective of molecular spectroscopy, the long-wavelength mid-IR is more suitable for identification of molecules, because the vibrational absorption spectrum of each molecule in the wavelength range of 6.7-25 µm, i.e. the fingerprint region, is distinctive [3, 4]. In strong-field physics, energy of the re-colliding electrons increases substantially with the driving wavelength, which enables the tunneling ionization even at moderate laser intensity, without causing an optical damage. This has triggered a number of studies on ultrafast electronics in molecules, solid materials as well as nano-structures [7-14]. In addition, long-wavelength mid-IR pulses have their photon energies far below the typical electronic interband resonances of bulk semiconductors. Therefore, phase-stable few-cycle long-wavelength mid-IR pulses could serve as a precisely adjustable bias for studies of ultrafast electronics [11]. With the demand for high-fidelity mid-IR spectroscopy and strong-field electron dynamics, high-energy few-cycle long-wavelength mid-IR lasers with stable carrier-envelop phase (CEP) are highly desired.

In the absence of broadband laser gain media in the mid-IR region, parametric down conversions such as optical parametric amplifier (OPA), optical parametric chirped-pulse amplifier (OPCPA) and difference-frequency generation (DFG) are commonly used to generate high-energy few-cycle mid-IR pulses. Non-oxide nonlinear crystals such as $AgGaS_2$ [15, 16], $CdSiP_2$ [17], and $ZnGeP_2$ [18-20] are employed to extend high-energy few-cycle pulses to the wavelength range of 4-11 µm. 8.5 µm, 150 µJ, few-cycle pulses are demonstrated by mixing the 1.8 µm pump and the 2.4 µm signal in an $AgGaS_2$-based DFG system [15]. 33 µJ, 0.88-cycle pulses covering 2.5-10 µm spectral range are generated by coherent synthesis of signal and idler pulses of a $CdSiP_2$ OPA, pumped by a 2 µm OPCPA [17]. The energy of a few-cycle $ZnGeP_2$-based OPCPA system at a central wavelength of 7 µm is boosted to half milijoule, with a powerful 2 µm cryogenic-cooled pump [18]. Recently, the wide bandgap nonlinear crystal $LiGaS_2$ has emerged. 7-11 µm few-cycle pulses with nano-joule-level pulse energy have been demonstrated in OPA and/or intrapulse DFG configurations based on $LiGaS_2$ crystal and pumped at the near-IR [21-24]. To pursue longer wavelength up to 20 µm, $AgGaSe_2$ [25, 26] and GaSe [27-29] nonlinear crystals have been employed in the DFG configuration, thanks to their longer transparent cutoff wavelength. In order to achieve broad phase-matching (PM) bandwidth and avoid parasitic two- or three-photon absorption, such light sources should ideally be pumped at ~2 µm wavelength [29]. Compared to $AgGaSe_2$, GaSe has higher damage threshold, larger nonlinear coefficient, larger birefringence (shown in Table. 1) and broader PM bandwidth pumped at ~2 μm wavelength [25-29]. These make GaSe a better candidate for generation of high-energy ultra-broadband few-cycle long-wavelength mid-IR pulses.

Table. 1 A comparison between GaSe and AgGaSe$_2$. Nonlinear coefficients ($d_{eff}$) are taken from SNLO for 8 μm idler.

| Crystals | Type-I,II $d_{eff}$ (pm/V) | Birefringence ($\Delta n$) | Damage threshold (TW/cm$^2$) | Transmission (μm) |
|---|---|---|---|---|
| GaSe | 57,55 | 0.34 [30] | 1.7 [28] | 0.65-18 [30] |
| AgGaSe$_2$ | 25,33 | -0.024 [30] | >0.2 [25] | 0.78-18 [30] |

In this letter, we report a GaSe-based mid-IR OPA driven by a ~2 μm source. The generated ultra-broadband idler pulse has a spectrum spanning from 4.2 to 16 μm, which supports a Fourier-transform-limited pulse width of ~19 fs (0.65 cycle), centered at 8.8 μm. The spectrum is much broader than the reported parametric sources based on other nonlinear crystals [15-26]. ~3.4 μJ idler pulse energy is obtained with a conversion efficiency of ~2%. We believe the generated multi-microjoule, few-cycle, ultra-broadband, long-wavelength mid-IR pulses would be impactful for both mid-IR molecular spectroscopy and ultrafast electron dynamics studies.

Type-I phase match is used in GaSe due to its broader PM bandwidth compared with Type-II phase match [29]. Fig. 1 shows the PM function $|sinc(\Delta kL/2)|$ with respect to the PM angle and the PM wavelength, where a 1-mm thick GaSe crystal and a pump wavelength of 2.15 μm are used for the calculation. Within 0.4 degree deviation of the PM angle, centered at ~11.1°, the PM bandwidth of the idler pulse ranging from 4 to 16 μm could be realized. It is worth mentioning that when a broadband pump source is used, even broader PM bandwidth can be obtained.

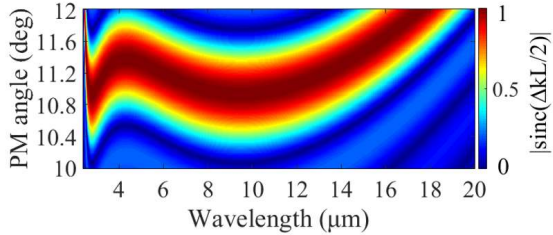

Fig. 1. The phase-matching (PM) function $|sinc(\Delta kL/2)|$ with respect to the PM angle and the PM wavelength, in a GaSe crystal with a length (*L*) of 1 mm, for the Type-I phase match, at a 2.15 μm pump wavelength.

The schematic of the mid-IR OPA is shown in Fig. 2(a). The pump source is a commercial multi-stage OPA system (TOPAS from Light Conversion, which is driven by a 5 mJ, 26 fs, 800 nm Ti: Sapphire laser system) with a 2.15 μm central wavelength, 420 μJ pulse energy, 51 fs pulse width, and 1 kHz repetition rate. A CaF$_2$ wedge placed at 22° with respect to the pump beam, functioning as a beam splitter, is employed to reflect ~10 μJ, 2.15 μm, *p*-polarized pump for the supercontinuum (SC) generation. It is rotated to *s*-polarization by a half-wave plate, and focused into a 6-mm thick BaF$_2$ by L$_1$ to generate signal pulses via SC. The generated SC is collimated by L$_2$ and resized to ~2.5 mm 1/$e^2$ diameter by a telescope comprised of L$_3$ and L$_4$. The transmitted 2.15 μm enters the pump line, which includes a delay line and a telescope (L$_5$ and L$_6$), to form a collimated pump beam with a comparable beam size as that of the signal beam.

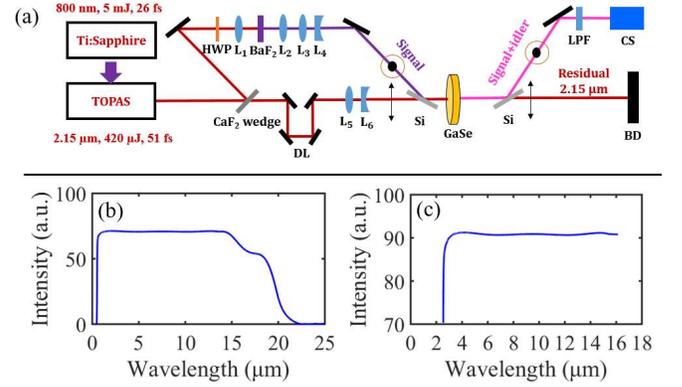

Fig. 2. (a) The schematic of the GaSe-based mid-IR optical parametric amplifier (OPA). The pump, the generated supercontinuum, and the amplified mid-IR pulses are shown in maroon, purple, and pink, respectively. (HWP-half wave plate, DL-delay line, LPF-long pass filter, BD-beam dump, CS-characterization setup, and L$_1$-L$_6$ are CaF$_2$ lenses.) The transmission curves of the uncoated ZnSe lens (b) and the mid-IR hollow-core fiber (c).

The collimated beam is then employed to pump a 1 mm-thick uncoated GaSe crystal with a Type-I phase match. As the GaSe crystal could be cleaved only along the (001) plane (z-cut, $\theta = 0°$), an internal PM angle of ~11.1° corresponding to an external angle of ~32° is introduced. This causes ~16% loss for the *p*-polarized pump. It is noted that the pump beam is routed to the crystal using multiple silver mirrors with a ~96% reflection for each. Taking into account all the losses, the maximum available pulse energy of the 2.15 μm pump is ~300 μJ, giving an estimated peak intensity of ~224 GW/cm$^2$. Two 200-μm thick silicon windows placed at a Brewster angle of 73.8° with respect to the pump beam are used as the beam combiner and beam splitter, respectively. This produces ~71% reflection for the *s*-polarized signal and idler beams. The generated idler beam is separated from the amplified signal beam using a long-pass filter (LPF) with a cut-off wavelength of 4.5 μm. A thermal sensitive power meter with a resolution of 1 μW (S401C, Thorlabs) and a calibrated mid-IR monochromator with a liquid-nitrogen-cooled HgCdTe (MCT) detector are used to characterize the mid-IR pulses. For the spectral measurement, an uncoated ZnSe lens with a near-flat transmission over 1-15 μm (Fig. 2(b)) and a dielectric-coated mid-IR hollow-core fiber with a 500-μm core diameter (HF500MW, OptoKnowledge) are employed to couple the mid-IR pulses into the monochromator. The hollow-core fiber has a near-flat transmission in the range of 3-16 μm (Fig. 2(c)).

The spectral and temporal profiles of the 2.15 μm pump from TOPAS are firstly characterized. The measured spectrum as shown in Fig. 3(a) has a full width at half maximum of ~220 nm which supports ~31 fs Fourier-transform-limited pulse width. The intensity autocorrelation shown in Fig. 3(b) reveals the pulse width of ~51 fs, assuming a Gaussian temporal profile. This agrees with the residual dispersion from the TOPAS optics, including about −400 fs$^2$ dispersion from the 3 mm fused silica beam splitter (reflecting 800 nm, and transmitting 2.15 μm). Fig. 3(c) shows the long-wavelength side of the SC spectrum (measured through a InF$_3$ fiber and a 2.4 μm LPF) which serves as the signal of the mid-IR OPA. It extends up to ~4.3 μm. With the knowledge of the long-wavelength edge of the signal, the calculated spectral edge on the short-wavelength side of the generated idler is <4.3 µm, which means the signal and idler overlap in the frequency domain. Thus a LPF with a 4.5 µm cut-off wavelength is employed to separate the generated idler from the amplified signal.

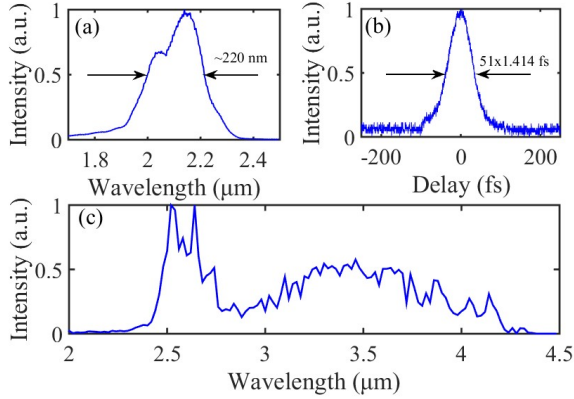

Fig. 3. (a) The measured pump spectrum from the TOPAS source with a ~420 µJ output energy. (b) The intensity autocorrelation trace of the pump pulses with the ~420 µJ output energy. (c) The supercontinuum generated from 6-mm-thick $BaF_2$, which serves as the signal of the mid-IR OPA, measured using 2.4 µm LPF and $InF_3$ fiber with 300-µm core diameter.

An ultra-broadband idler spectrum is obtained by carefully optimizing the crystal angle, temporal and spatial overlaps between the pump and the signal. The idler spectrum spans from 4.2 to 16 µm, with the -3 dB bandwidth ranging from 4.5 to 13.3 µm, shown in Fig. 4(a). It should be noted that the 4.5 µm LPF is responsible for the steep edge at ~4.6 µm. The distance from the OPA output to the detector of the monochromator is ~3 m, thus some fine structures caused by the atmospheric absorption appear in the spectrum. Similar absorption structures have been reported in the broadband mid-IR systems [18, 31]. In addition, the SC components in the range of 2.5-3 µm are located in the water absorption window. These absorption structures are transferred to the 7-15 µm idler spectrum during the parametric amplification, which also contributes to the fine structures of the idler spectrum. A vacuum or nobel gas purging chamber could be used to remove these spectral structures [28].

The OPA process is further investigated using a (2+1)-dimensional numerical simulation [32]. Taking into account the high pump intensity and the large nonlinear refractive index of GaSe ($n_2 = 450 \times 10^{-1}$ $cm^2/W$) [33], the self-phase modulation and self-focusing are included in the calculations, which are found crucial for reproducing the measured spectrum. A 2.15 µm, 300 µJ, 50 fs chirped Gaussian pump pulse with a spectrum spanning from 1.8 to 2.3 µm, and a 2.84 µm, 3 nJ, 50 fs chirped signal with a spectral shape similar to the measured SC covering 2.3 to 4.3 µm are employed in the simulation to mimic the experimental conditions. The PM angle is set to $11.15°$. The simulated spectra are shown in Fig. 4(a) and (b), which qualitatively match with the measured idler and amplified signal spectra. The slight mismatch for idler spectrum at > 13 µm is due to the crystal absorption [34].

In order to further verify the accuracy of the measured idler spectrum, we integrate the measured spectrum and calculate the energy distribution in different spectral bands, i.e. 4.2-7.3 µm, 7.3-11.7 µm, and 11.7-16 µm, based on the measured idler pulse energy. We compare the integrated pulse energies with the measured, as shown in Fig. 4(c). The good agreement between the integrated and measured pulse energy gives a direct evidence for the accuracy of the spectrum measurement.

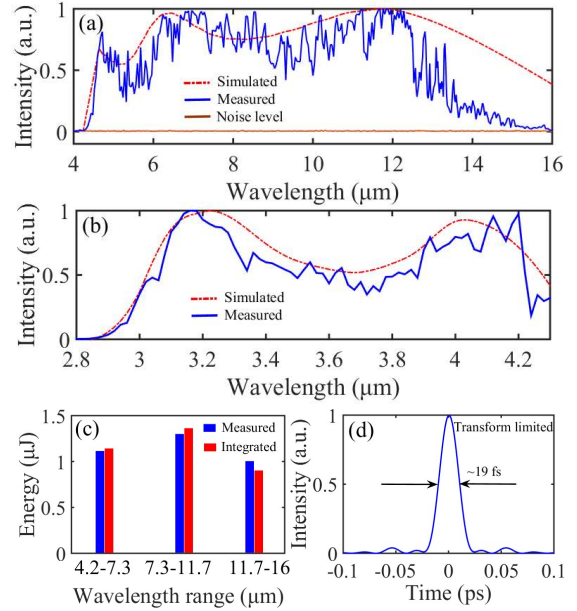

Fig. 4. The measured (solid blue, 20 nm resolution) and simulated (dashed red) spectra of the idler pulses (a) and the amplified signal pulses (b). Note that the measured spectra are calibrated by taking account of the response curves of the grating, MCT, LPF and the hollow-core fiber. The simulated spectra include the edge responses of the hollow-core fiber and LPF. (c) The measured and integrated energy distribution in different bands of the idler wavelength. (d) The calculated Fourier-limited pulse width based on the measured idler spectrum.

Based on the idler spectrum, the calculated transform-limited pulse width is ~19 fs (Fig. 4(d)), which corresponds to 0.65-cycle pulse width, centered at 8.8 µm. The actual pulse width could be broadened to few cycles due to the dispersion from the 1 mm thick 4.5 µm LPF (Ge substrate) and the intrinsic dispersion from the OPA. It is worth noting that as the signal and the pump are from the same TOPAS pump system, the CEP fluctuation in the idler pulse is self-canceled through the DFG nature of the OPA [35, 36]. Recently, stable CEP of the idler pulse from a passively CEP stable mid-IR OPA has been experimentally confirmed [17]. Thus, we, to the large extent, believe that the CEP for this system is stable. Its CEP stability could be degraded by some factors such as the intensity-to-phase noise of the pump, and the temporal jitters between the pump and the signal [35, 36].

The pulse energy of the generated idler is measured as ~3.4 µJ behind a 4.5 µm LPF, with ~300 µJ pump energy. Considering the 85% transmission of the LPF and ~71% reflection from the silicon beam splitter for the idler pulses, ~2% conversion efficiency from the pump to the idler is achieved. For this OPA system, the measured OPA output by blocking the signal is ~0.12 µJ, and the energy fluctuation of the generated idler pulses is 2.8% rms for a measurement duration of 30 minutes. In addition, the idler pulse energies with different LPFs as a function of the pump energy are measured, shown in Fig. 5. No saturation occurs at the maximum pump energy, which implies further energy scaling up is possible with higher pump energy. The idler beam profile behind the 4.5 μm LPF are also measured, exhibiting a good Gaussian profile shown in the inset of Fig. 5.

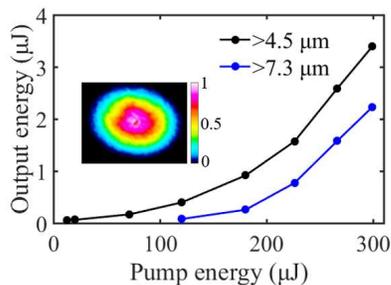

Fig. 5. The dependence of the idler pulse energy on the pump energy behind the 4.5 and 7.3 μm LPFs. The inset is the beam profile of the idler pulse behind the 4.5 μm LPF at the output energy of ~3.4 μJ.

In conclusions, 3.4 μJ ultra-broadband mid-IR pulses with a spectrum spanning from 4.5 to 13.3 μm at –3 dB and from 4.2 to 16 μm in the full spectral range are generated from a SC seeded GaSe-based OPA system pumped at ~2 μm. To the best of our knowledge, this is the broadest –3dB spectrum ever obtained by OPA systems in this wavelength region. With the rapid progress of ~2 μm Ho:YLF or Ho:YAG CPA systems [18-20, 37], the energy scaling up of the ultra-broadband mid-IR pulses is expected through the development of the GaSe-based OPCPA systems.

**Acknowledgement.** We acknowledge the financial support from SERC (Grant No.1426500050, and 1426500051) from the Agency for Science, Technology and Research (A*STAR), Singapore.